\documentclass[12pt]{article}
\usepackage{amsfonts,latexsym,fullpage,amsmath,latexsym,amssymb,epsf}

\date{May 16, 2003}

\newcommand{\cB}{{\cal B}}

\newcommand{\ket}[1]{|#1\rangle}
\newcommand{\bra}[1]{\langle #1|}

\newcommand{\beq}{\begin{equation}}
\newcommand{\eeq}{\end{equation}}
\newcommand{\beqy}{\begin{eqnarray}}
\newcommand{\eeqy}{\end{eqnarray}}

\newtheorem{Definition}{Definition}

\newtheorem{Theorem}{Theorem}

\newenvironment{Proof}{{\it Proof: \,}}{$\Box$ \vspace{0.3cm}}

\newenvironment{Definition*}{{\bf Definition}}{}

\def\C{{\mathbb{C}}}

\newcommand{\B}{{\cal B}}
\newcommand{\cH}{{\cal H}}

\begin{document}

\title{Two QCMA-complete problems}

\author{Pawel Wocjan\thanks{e-mail: 
\{wocjan,janzing\}@ira.uka.de}, Dominik Janzing, and Thomas
Beth \\ \small Institut f{\"u}r Algorithmen und Kognitive Systeme,
Universit{\"a}t Karlsruhe,\\[-1ex] \small Am Fasanengarten 5,
D-76\,131 Karlsruhe, Germany}

\maketitle

\abstract{QMA and QCMA are possible quantum analogues 
of the complexity class 
NP.  In QCMA the verifier is a quantum program and the proof is classical.
In contrast, in QMA the proof is also a quantum state.

We show that two known QMA-complete problems can be modified to 
QCMA-complete problems in a natural way:

(1) Deciding whether a $3$-local Hamiltonian has low energy states
(with energy smaller than a given value) that can be prepared
with at most $k$ elementary gates is QCMA-complete, whereas
it is QMA-complete when the restriction on the complexity of preparation
is dropped.

(2) Deciding whether a (classically described) quantum circuit
acts almost as the identity on {\it all basis states} is QCMA-complete. 
It is QMA-complete to decide whether it acts on {\it all states}
almost as the identity.}

\section{Introduction}

The complexity class QCMA is the class of decision problems for which
a ``yes'' answer can be verified by a quantum computer with access to
a {\em classical} proof. It contains MA \cite{Babai}, 
and is contained in QMA \cite{KitaevShen}. The
computer is restricted to be classical for MA and the proof is allowed
to be quantum for QMA.

More explicitly, QMA problems read as follows:
Given a quantum circuit $U$ that acts on a quantum register
consisting of $n+m$ qubits where $m$ qubits (the ``ancillas'')
are initialized to the state $|0\dots 0\rangle$,
decide whether there is a state vector $|\psi\rangle$ on the remaining
$n$ qubits (the ``input register'') such that after the implementation of
$U$ a measurement of the first qubit yields ``1'' with high probability.
We say, the circuit has accepted the input state $|\psi\rangle$.

For QCMA, the problem is to decide whether there is a {\it basis
state} $|y\rangle$ 
 on $n$ qubits that is accepted with high probability. Then the
classical proof consists merely of the number $0\leq y < 2^n$ 
of the basis state.

Let us recall the formal definition of QCMA \cite{dorit}.
In the following we denote the vector space $\C^2$ by $\cB$
and the length of any binary string $y\in \{0,1\}^*$ by $|y|$.

\begin{Definition}[QCMA] \label{QCMA}${}$\\
Fix $\epsilon=\epsilon(|x|)$ such that $2^{-\Omega(|x|)} \leq \epsilon
\leq 1/3$.  Then a language $L$ is in QCMA if for every classical input
$x \in \{0,1\}^*$ one can efficiently generate (by classical
precomputation) a quantum circuit $U_x$ (``verifier'') consisting of
at most $p(|x|)$ elementary gates for an appropriate polynomial $p$
such that $U_x$ acts on the Hilbert space
\[
\cH:= \B^{\otimes {n_x}} \otimes \B^{\otimes m_x}\,,
\]
where $n_x$ and $m_x$ grow at most polynomially in $|x|$. The first
part is the input register and the second is the ancilla register.
Furthermore $U_x$ has the property that
\begin{enumerate}
\item If $x\in L$ there exists a classical string
$y\in\{0,1\}^{n_x}$ such that the corresponding computational basis
state $\ket{y}$ is accepted by the circuit with high probability,
i.e.,
\[\exists\, y\in \{0,1\}^{n_x}\,,\quad
tr(U_x\,(\ket{y}\bra{y}\otimes\ket{0\ldots 0}\bra{0\ldots 0})\,U^\dagger_x\,
P_1) \geq 1-\epsilon\,,
\]
where $P_1$ is the projection corresponding to the measurement ``Is
the first qubit in state $1$?''.
\item If  $x\not\in L$ all computational basis states are rejected
with high probability, i.e.,
\[
\forall\, y\in \{0,1\}^{n_x}\,,\quad
tr(U_x\,(\ket{y}\bra{y}\otimes\ket{0\ldots 0}\bra{0\ldots 0})\,U_x^\dagger\, 
P_1) \leq \epsilon\,.
\]
\end{enumerate}
\end{Definition}

\section{Identity check on basis states}

In a recent paper \cite{IdentityQMA} 
we stated the problem ``identity check''.
The task  is to decide whether a (classically described) quantum circuit $U$
is almost equivalent to the identity in the sense that there is a global 
phase $\phi$ such that the operator norm $\|U-\exp(i\phi) {\bf 1}\|$
is close to zero. This problem arises naturally in the design
of quantum circuits: Given a quantum circuit $U_l\cdots U_1$:
Decide whether another sequence of elementary gates
$V_k\cdots V_1$ 
implements almost the same unitary transformation, i.e.,
whether 
\[
U_l\cdots U_1 V_1^\dagger \cdots V_k^\dagger
\]
is almost equivalent to the identity.

But also a weaker definition of equivalence is natural.
Usually, quantum algorithms start with classical input 
(basis states as input) and end with measurements in the 
computational basis to obtain the classical output.
In this context one does not care whether two circuits agree on all
states, it is only relevant whether they agree on the basis states.
Below we shall show that this problem is QCMA-complete.

But what makes the difference between the original requirement and the weaker
formulation? First it is clear that a unitary operator that
maps every basis state $|x\rangle \langle x|$ on itself may give 
different phases to different basis states.
But one can see easily that this {\it does not} make the difference
between QCMA and QMA (in case these classes are indeed different):
The statement that a quantum circuit gives different phases to different
basis vectors has still a classical proof. It is given by two numbers
of basis states with non-negligible phase difference. The verifier can check
the phase difference efficiently 
by quantum phase estimation \cite{ClevePhase} 
(compare also \cite{IdentityQMA}). 

What can possibly make the difference between QMA and QCMA is the fact 
that there exist unitary transformations $U$ that have large norm distance
to all trivial transformations $\exp(i\phi) {\bf 1}$ even though 
the distance between $U|x\rangle$ and $|x\rangle$ 
is exponentially small on all basis
states $|x\rangle$.
Let $U=H D H$, where $H$ is the Hadamard transformation on $n$ qubits
and $D={\rm diag}(-1,1,1,\ldots,1)$ is a controlled phase shift on the
first qubit. The norm distance $\|{\bf 1}-D\|$ is $2$. But for all
computational basis states $\ket{y}$ we have
\[
\|({\bf 1}-H D H)\ket{y}\| = \|H({\bf 1}-D)H\ket{y}\| 
=
\|(2/2^n)\,\sum_{\tilde{y}} \ket{\tilde{y}}\| 
= 
2/2^{n/2}
\]
since $H\,{\rm diag}(1,0,0,\ldots,0)\,H$ is the all-one-matrix.

Let us define the problem ``identity check on basis states''.
\begin{Definition}[Identity check on basis states]\label{def:QCMA}${}$\\
Let $x$ be a classical description of a quantum circuit $Z_x$ on $n$ 
qubits. Given
the promise that
\begin{enumerate}
\item either there is a binary string $z$ such that
\[
|\bra{z} Z_x \ket{z}|^2\le 1- \mu\,,
\]
i.e., $Z_x$ does not act as the identity on the basis states,
\item
or for all binary strings $z$
\[
|\bra{z} Z_x \ket{z}|^2\ge 1-\delta \,,
\]
i.e., $Z_x$ acts ``almost''  as the identity on the basis states,
\end{enumerate}
where $\mu-\delta \ge 1/poly(|x|)$.
Decide which case is true.
\end{Definition}
It is easily seen that this problem is contained in QCMA
since the proof for case 1 is given by a string that describes the
basis state $|z\rangle$. 
Then we perform the quantum circuit $Z_x$. 
An $n$-fold controlled-NOT can be used to flip an additional ancilla
qubit if and only if the output is $|y\rangle$. The additional ancilla
is the output qubit of the verifier. 

QCMA-completeness of this problem can be proved in strong
analogy to the proof for  QMA-completeness of ``identity check''. 
Let $U$ be
a quantum circuit as in Definition~\ref{def:QCMA}. We
construct the quantum circuit $Z$ that uses $U$ and $U^\dagger$ as 
subroutines.

Let $R$ be the rotation
\[
\left(
\begin{array}{cc}
\cos(\varphi) & -\sin(\varphi) \\
\sin(\varphi) & \cos(\varphi)
\end{array}
\right)\,,
\]
with $0<\varphi<\pi/2$
and $R_1$ be the rotation $R$ 
 controlled by the $m$ ancilla qubits corresponding
to $U$. 
$R_1$ is implemented if and only if the ancillas are correctly
initialized in the state $|0\dots 0\rangle$.
(In \cite{IdentityQMA} we have used controlled phase shift which was diagonal
in the computational basis).
Let $R_2$ be the same rotation $R$ controlled by the output qubit
of $U$. The whole circuit $Z:=U^\dagger R_2 U R_1$ 
is shown in Fig.~1. 

\begin{figure}
\centerline{
\epsfbox[0 0 208 155]{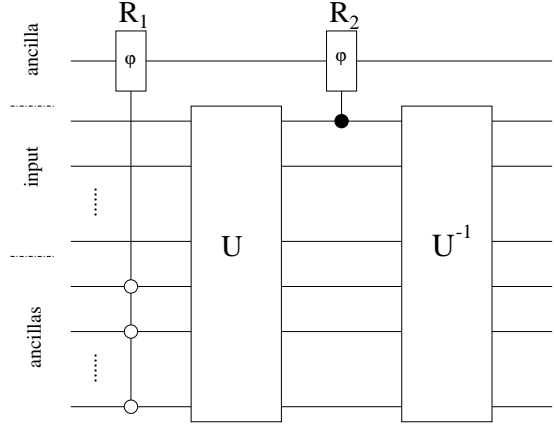}
}
\caption{The circuit $Z$ acts on all basis states almost as
the identity if and only if no basis state is likely to be accepted by $U$.} 
\end{figure}

The following theorem shows that the problem of deciding whether
there are basis states that are likely to be accepted by $U$ 
can be reduced to identity check on basis states.

\begin{Theorem}
Let $U$ be a quantum circuit on $\B^{n+m}$ with the promise that one
of the two cases in Definition~\ref{def:QCMA} is true. Then the
following statement holds for the corresponding circuit $Z$:

If case $1$ is true then there is a binary string $z$ such that
\[
|\bra{z}Z\ket{z}|^2 \le (\cos(2\varphi)+\sqrt{\epsilon})^2\,,
\]
where $\ket{z}=\ket{0}\otimes\ket{y}\ket{00\cdots 0}$ and $y$ is the
accepting classical witness for the the circuit $U$.

If case $2$ is true then for all binary strings $z$ we have
\[
|\bra{z}Z\ket{z}|^2 \ge (\cos(\varphi) - 2\sqrt{\epsilon})^2\,.
\] 
\end{Theorem}
\begin{Proof}
The proof is in strong analogy to the proof of Theorem 1 in 
\cite{IdentityQMA}. The important difference is that no superpositions
between states with correctly and wrongly initialized ancillas
have to be considered. Therefore the bounds are easier to derive.
 
Consider case~1. Let $\ket{y}$ be a binary string that is accepted by
$U$ with high probability (we drop the subscripts for fixed $x$). We
consider the binary string
$\ket{z}:=\ket{0}\otimes\ket{y}\otimes\ket{00\cdots 0}$ to show 
that $Z$ is
``far'' from the identity on the basis states.

\begin{eqnarray*}
Z\ket{z} & = & U^\dagger R_2 U R_1\ket{z} \\
& = &
U^\dagger R_2 U (\cos \varphi) |0\rangle +\sin \varphi |1\rangle)
\otimes |y\rangle \otimes |0\dots 0\rangle\\ &=&
U^\dagger R_2  (\cos \varphi) |0\rangle +\sin \varphi |1\rangle)
\otimes ( c_1 |1\rangle \otimes |\psi_1\rangle + c_0 |0\rangle \otimes
|\psi_0\rangle )
\end{eqnarray*}
Due to the high probability of acceptance we have $|c_0|\leq \sqrt{\epsilon}$.
Now we consider only the term with $c_1$ and obtain
\begin{eqnarray}\label{vector}
&&U^\dagger R_2  (\cos \varphi) |0\rangle +\sin \varphi |1\rangle)
\otimes c_1 |1\rangle \otimes |\psi_1\rangle  \nonumber \\&=&
(\cos (2\varphi)) |0\rangle +\sin (2\varphi) |1\rangle)
\otimes c_1 U (|1\rangle \otimes |\psi_1\rangle)\,.
\end{eqnarray}
The first component is the single ancilla on which 
the rotation $R$ is performed, the second  component is the output of $U$ 
and the third tensor component is the remaining part of the register
where $U$ acts on.

The overlap between the initial vector $|z\rangle$ and
the vector of eq.~(\ref{vector}) is at most $|c_1|\cos(2\varphi)$.
Taking into account the length of the neglected vector  
we obtain
\[
|\langle z| Z|z\rangle |\leq \cos(2\varphi) + \sqrt{\epsilon}\,.
\]

Consider case~2. Let $z$ be a string such that the bits
corresponding to ancillas of $U$ are all set to $0$.
Let $P_1$ as in Definition~\ref{QCMA} be the
projection onto the state $|1\rangle$ of the output qubit corresponding
to $U$.
Note that $R_2 ({\bf 1} - P_1)= {\bf 1} -P_1$. 
 Then we have
\begin{eqnarray*}
|\bra{z}Z\ket{z}| 
& = & 
|\bra{z}U^\dagger R_2 U R_1\ket{z}| \\
& = &
|\bra{z}U^\dagger R_2(P_1 + {\bf 1} - P_1) U R_1\ket{z}| \\
& = &
|\bra{z} U^\dagger R_2 P_1 U R_1 + R_1 - U^\dagger P_1 U R_1\ket{z}| \\
& \ge &
|\bra{z}R_1\ket{z}| - |\bra{z}U^\dagger R_2 P_1 U R_1\ket{z}| -
|\bra{z}U^\dagger P_1 U R_1\ket{z}| \\
& \ge &
\cos\varphi - 2\sqrt{\epsilon}\,.
\end{eqnarray*}
The latter inequality is due to the fact that the length
of the vector $P_1UR_1|z\rangle$ is at most $\sqrt{\epsilon}$ due to
the small probability of acceptance.

Let $z$ be a string such that the bits corresponding to the ancillas
of $U$ are not all set to $0$. Then we have
\begin{eqnarray*}
|\bra{z} U^\dagger R_2 U R_1\ket{z}| 
& = &
|\bra{z} U^\dagger R_2 U\ket{z}| \\
& \ge &
\cos(\varphi) \,.
\end{eqnarray*}
This can be seen by writing $\ket{z}$ as
$\ket{\Psi_{-\varphi}}\oplus\ket{\Psi_0}\oplus\ket{\Psi_{\varphi}}$,
where $\ket{\Psi_{-\varphi}},\ket{\Psi_0}$ and $\ket{\Psi_{\varphi}}$
are vectors in the eigenspaces of $U^\dagger R_2 U$ corresponding to
the eigenvalues $e^{-i\varphi},1$ and $e^{i\varphi}$, respectively. 
Therefore,
we have
\[
|\bra{z}U^\dagger R_2 U\ket{z}| = |p e^{-i\varphi}+qe^{i\varphi}+r|
\]
with $p:= \|\,\ket{\Psi_{-\varphi}}\|^2$, $q:=\|\,\ket{\Psi_0}\|^2$
and $r:=\|\,\ket{\Psi_{\varphi}}\|^2$. By elementary geometry
the shortest vector in the convex span of the complex values 
$e^{-i\varphi},1,e^{i\varphi}$ has length $\cos(\varphi)$.
This completes the proof.
\end{Proof}

\section{Low energy states with low complexity}

The  problem of determining the ground state energy and spectral gaps 
of many-particle
systems is a highly non-trivial task. 
In \cite{KempeRegev} (based on results in \cite{KitaevShen}) 
it was shown that
even the following instance is QMA-complete:
Given a $3$-local Hamiltonian, i.e., a selfadjoint operator
on $n$-qubits which is a sum of operators that act on only $3$ qubits.
Furthermore, let the promise be given that either all eigenvalues
of $H$ are at least $b$ or there exists an energy value
smaller or equal to $a$.
Decide which statement of both is true.

But there is a slightly different question which is interesting as well:
It is to decide whether there exist states with energy at most $a$ 
which are {\it simple} to prepare. 
Here simplicity will be defined by the number of required elementary gate
operations. 
This problem arises naturally when one
addresses the question whether extremely efficient 
cooling mechanisms could prepare states that are difficult 
to obtain with a reasonable number of quantum gates
\cite{TerhalDiVin,AharonovState}.

Note that it is not clear that it should be  easier to decide whether there
exist low-complexity states  with low energy than to decide whether 
low energy states (with energy  at most $a$)
do exist at all. In special instances one may possibly  
have arguments showing
that low eigenvalues exist although one has no idea how to prepare them.
However, here we show that the {\it problem class} of deciding
whether a general $3$-local Hamiltonian has low-energy states includes
the problem of deciding whether there are low complexity low energy states.
This follows from the fact that the first problem class is QMA and the latter
one is QCMA.

Now we state the considered problem class 
formally.

\begin{Definition}[Low complexity low energy states]${}$\\
Given a 
$3$-local Hamiltonian $H$ on $n$ qubits and  real numbers $a,b$ with
$b-a \geq 1/poly (n)$. 
Let the set of elementary gates be the Shor basis \cite{Boykin}.
Then the problem ``low complexity low energy states''
is to decide which one of the following cases is true given the promise
that either of two holds:

\begin{enumerate}

\item
There is a sequence of $k$ elementary gates
such that 
\[
|\psi\rangle := V|0\dots 0\rangle 
\]
is a state with energy less than a, i.e.
\[
\langle \psi | H |\psi \rangle \leq a\,.
\]

\item
All quantum circuits $V$ that consist of at most $k$ gates
can only prepare states $|\psi\rangle$ 
with energy at least $b$, i.e.,
\[
\langle \psi | H |\psi \rangle \geq b\,.
\]

\end{enumerate}

\end{Definition}

We obtain the following theorem:

\begin{Theorem}
The problem ``low complexity
low energy states'' is QCMA-complete.
\end{Theorem}

\begin{Proof}
It is easy to see that the problem is in QCMA: the witness is a classical
string describing the preparation procedure. The fact that 
$|\psi\rangle$  is indeed 
a state with energy not greater than $a$ can be checked as in 
\cite{KitaevShen}.

Now we  show that the problem encompasses QCMA.  We consider a quantum
circuit
$U$ and the task is to decide whether there is a basis state that is
accepted with high probability. 
Let $n$ be the number of qubits of the input 
register and $m$ be the number of ancilla qubits. 
Then we construct a circuit $\tilde{U}$ with $n$ input 
qubits and $n+m$ ancillas
as follows: $n$ C-NOT gates copy the input to the $n$ additional ancillas.
It is easy to see that $\tilde{U}$ has a state that is accepted 
with high probability if and only
if $U$ accepts a basis state with high probability.

Now we use the construction of \cite{KempeRegev} and obtain a $3$-local 
Hamiltonian
$H$ associated with $\tilde{U}$. Let $L$ be the number of gates of 
$\tilde{U}$. The exact form of $H$ is not important here, we only rephrase
the following four results of \cite{KempeRegev,KitaevShen}
which are necessary for our proof:

\begin{enumerate}

\item The Hamiltonian corresponding to $H$ acts on 
$\tilde{n}+\tilde{m}+L$ qubits where $\tilde{n}+\tilde{m}$ is the size of
 the rgister where $\tilde{U}$ acts on (the input is an $\tilde{n}$-qubit
state and the ancilla register of $\tilde{U}$ consists of $\tilde{m}$ qubits.
The additional register with size $L$ is a so-called ``clock'' register.
Its role is not relevant here.

\item Whenever the circuit $\tilde{U}$ 
rejects all states with probability at least
$1-\epsilon$ there is no eigenvalue of $H$ smaller or equal to  $c/L^3$ 
for an appropriate constant $c$.

\item If the state $|\psi \rangle$ is accepted by  $\tilde{U}$ 
with probability $1-\epsilon$ the state
\[
|\eta\rangle:= \frac{1}{\sqrt{L+1}}
\sum_{j=0}^L U_j\cdots U_1(|\psi\rangle \otimes |0\dots 0\rangle)
\otimes |
2^j-1\rangle
\]
is a low energy state, i.e.,
\[
\langle \eta | H |\eta \rangle \leq \frac{\epsilon}{L+1}\,.
\]

\item In case that the ``confidence value'' $\epsilon$ is too large
such that $\epsilon/(L+1) \geq c/L^3$ 
or the gap between both values is too small,
one may use probabbility amplification
\cite{KitaevShen} and define a circuit 
$\tilde{U}'$ given by many parallel implementations of $\tilde{U}$ 
with majority vote in the end such the promise in Definition~\ref{QCMA}
holds for a smaller value $\epsilon'$.
Then the method of \cite{KempeRegev}
is applied to obtain a Hamiltonian
$H$ corresponding to $\tilde{U}'$ such that 
$\epsilon/(L+1)$ is sufficiently smaller than $c/L^3$.
\end{enumerate}

Consider the case that there is no basis state input of $U$ that is
accepted with probability greater than $\epsilon$.
Then there is also no input 
state at all that is accepted by $\tilde{U}$ 
with probability greater than $\epsilon$.
As rephrased above,
there is no eigenvalue of $H$ smaller than $c/L^3$.

Consider the case that there is a basis state $|x\rangle$ 
that is accepted by $U$ 
with probability at least $1-\epsilon$. It is accepted by
$\tilde{U}$ with the same probability.
Then the state  $|\eta\rangle$ defined above is a low energy state.
It can be prepared efficiently,
i.e. there is a polynomial $p$ such that $|\eta\rangle$ 
can be obtained by $p(|x|)$ elementary gates.
We omit technical details but it is not difficult to show that
the superposition 
\[
\frac{1}{\sqrt{L+1}}\sum_{j=0}^L |2^j-1\rangle
\]
can be prepared efficiently. By applying $L+1$-fold controlled
$U_j$-gates one obtains the state $|\eta\rangle$. 

The question whether there is
a low energy state that can be prepared with at most $p(|x|)$ elementary 
gates
is hence equivalent to  the question whether there is a basis state that
is accepted by $U$.
\end{Proof}

\section{Remark on other  problems in QCMA}

To find simple procedures for preparing certain 
entangled multi-particle states from unentangled initial states 
is an interesting question of quantum information theory. 
Once one has found a procedure that prepares a desired state
$|\psi\rangle$ from the state $|0\dots 0\rangle$, for instance,
one may want to know whether there is also a simpler way to prepare
$|\psi\rangle$.

Hence the following type of problems seems natural:
Given a classical description of a quantum circuit $U$, decide whether
there is also a simpler preparation procedure for $|\psi \rangle:= 
U|0\dots 0\rangle$ in the following sense:
 
\begin{enumerate}

\item 
Given an elementary set of universal quantum gates. 
Decide whether there exists a 
quantum circuit $V$ consisting of at most $k$ gates preparing
almost the same state (norm difference at most $1- \delta$)  
or all states prepared using at most $k$ 
gates have at least the norm distance $1-\mu$ from $|\psi\rangle$

\item
Let  $l$ and $T$ be given, 
decide whether there is a $l$-local Hamiltonian preparing $|\psi\rangle$
approximatively by its autonomous evolution within the time $T$, i.e.,
\[
\exp(-iHt) |0\dots 0\rangle
\]
is almost the same state as $|\psi\rangle$ for an appropriate
value  $t\leq T$.

\item
Consider the control-theoretic setting as, for instance, the one appearing
in NMR-experiments \cite{KBG}:
Given a pair-interaction Hamiltonian $H$ and a maximal running time 
$T$.
Let 
a state $\ket{\Psi}$ 
be specified 
by a quantum circuit as above,
decide if it is possible to intersperse the natural 
time-evolution
by at most $k$ 
fast local operations (i.e. one-qubit rotations) 
such that the resulting
unitary prepares the desired state  such that the maximal running time
$T$ is not exceeded.

\end{enumerate}

These types of problems are clearly in QCMA 
when the desired accuracies are defined as in Definition~\ref{def:QCMA} 
since the proof consists
of a classical description of the preparation procedure (i.e. the
gate sequence, the Hamiltonian or the control sequence).
The verifier can check that the procedure does indeed prepare the desired 
state by 
simulating the described preparation procedure on a quantum computer and
applying $U^\dagger$. Then he measures the obtained state
in the computational basis.

We do not know whether these problems are contained in any lower 
complexity class.

This work was supported by the BMBF-project 01/BB01B.


\begin{thebibliography}{10}

\bibitem{Babai}
L.~Babai and S.~Moran.
\newblock Arthur-Merlin games: a randomized proof system, and a hierarchy of
  complexity classes.
\newblock {\em Journal of Computer and System Sciences}, 36(2):254--276, 1988.

\bibitem{KitaevShen}
A.~Kitaev, A.~Shen, and M.~Vyalyi.
\newblock {\em Classical and Quantum Computation}, volume~47.
\newblock Am. Math. Soc., Providence, Rhode Island, 2002.

\bibitem{dorit}
D.~Aharonov and T.~Naveh.
\newblock Quantum NP - a survey.
\newblock {\em quant-ph/0210077}.

\bibitem{IdentityQMA}
D.~Janzing, P.~Wocjan, and T.~Beth.
\newblock ``Identity check'' is QMA-complete.
\newblock {\em quant-ph/0305050}.

\bibitem{ClevePhase}
R.~Cleve, A.~Ekert, C.~Macchiavello, and M.~Mosca.
\newblock Quantum algorithms revisited.
\newblock {\em Proc. Roy. Soc. London A}, 454:339--354, 1998.


\bibitem{KempeRegev}
J.~Kempe and O.~Regev.
\newblock 3-local Hamiltonian is QMA-complete.
\newblock {\em quant-ph/0302079}, 2003.

\bibitem{TerhalDiVin}
B.~Terhal and D.~DiVincenzo.
\newblock The problem of equilibration and the computation of correlation
  functions on a quantum computer.
\newblock {\em quant-ph/9810063}.

\bibitem{AharonovState}
D.~Aharonov and A.~Ta-Shma 
\newblock Adiabatic quantum state generation and statistical zero knowledge.
\newblock {\em quant-ph/0301023}.

\bibitem{Boykin}
P.~Boykin, T.~Mor, M.~Pulver, V.~Roychowdhury, and F.~Vatan.
\newblock On universal and fault-tolerant quantum computing: 
A novel basis and
  a new constructive proof of universality for Shor's basis.
\newblock {\em Prooceedings of the 40th Annual Symposium on foundations of
  Computer Science}, pages 486--494, 1999.

\bibitem{KBG}
N.~Khaneja, R.~Brockett, and S.~Glaser
\newblock Time optimal control in spin systems.
\newblock {\em Phys. Rev. A}, 63(3):032308--1--13, 2001.

\end{thebibliography}
\end{document}